\documentclass[twocolumn,superscriptaddress,showpacs,preprintnumbers,nofootinbib,amsmath,amssymb,floatfix,aps]{revtex4-1}
\usepackage{dcolumn}% Align table columns on decimal point
\usepackage{bm}% bold math
\usepackage{color}

\usepackage{graphicx}% Include figure files
\usepackage{amsmath}
\usepackage{amssymb}
\usepackage{pstricks}

\def\gsim{\buildrel > \over {_{\sim}}}

\def\beq{\begin{equation}}
\def\eeq{\end{equation}}
\def\be{\begin{eqnarray}}
\def\ee{\end{eqnarray}}
\def\ra{\rangle}

\begin{document}
%\preprint{APS/123-QED}
\title{Comparison of the electromagnetic responses of $^{12}$C obtained from the Green's function Monte Carlo and spectral function approaches}
%%%%%%%%%%%%%%%%%%%%%%%%%%%%%%%%%%%%%%%%%%%%%%%%%%%%%%%%%%%%%%%%%%%%%%%%%
\author{Noemi Rocco}
\affiliation{INFN and Department of Physics, ``Sapienza'' University, I-00185 Roma, Italy}
\affiliation{Instituto de Fisica Corpuscular (IFIC), Centro Mixto CSIC-Universidad de Valencia,
Institutos de Investigacion de Paterna, E-46071 Valencia, Spain}
\author{Alessandro Lovato}
\affiliation{Physics Division, Argonne National Laboratory, Argonne, Illinois 60439, USA}
\author{Omar Benhar}
\affiliation{INFN and Department of Physics, ``Sapienza'' University, I-00185 Roma, Italy}
%\affiliation{Center for Neutrino Physics, Virginia Tech, Blacksburg, Virginia 24061, USA} 

%%%%%%%%%%%%%%%%%%%%%%%%%%%%%%%%%%%%%%%%%%%%%%%%%%%%%%%%%%%%%%%%%%%%%%%%%
\date{\today}
%%%%%%%%%%%%%%%%%%%%%%%%%%%%%%%%%%%%%%%%%%%%%%%%%%%%%%%%%%%%%%%%%%%%%%%%%
\begin{abstract}
The electromagnetic responses of carbon obtained from the  Green's function Monte Carlo and spectral function approaches using the same dynamical input  are compared in the kinematical 
region corresponding to momentum transfer in  the range 300\textendash570 MeV. The results of our analysis, aimed at  pinning down the limits of applicability of 
the approximations involved in the two schemes, indicate that the factorization {\em ansatz} underlying the spectral function formalism  provides remarkably 
accurate results down to momentum transfer as low as 300 MeV. On the other hand,  it appears that at 570 MeV relativistic corrections to the electromagnetic current not included 
in the Monte Carlo calculations may play a significant role in the transverse channel.
 %{\blue
%We present a systematic comparison between the Green's function Monte Carlo and the spectral function approaches for the calculation of the electromagnetic longitudinal and transverse response functions of $^{12}$C. The role of final state interactions and relativistic effects, both in the transition operators and in the final state, is elucidated by analyzing three kinematical setups. The agreement between the two methods, based on the same nuclear dynamical model, is remarkably good. Our results corroborate the reliability of the factorization {\em ansatz} in obtaining reliable predictions of the lepton-nucleus cross section in the kinematical region relevant for neutrino-oscillation experiments.
%}
\end{abstract}
\maketitle
%%%%%%%%%%%%%%%%%%%%%%%%%%%%%%%%%%%%%%%%%%%%%%%%%%%%%%%%%%%%%%%%%%%%%%%%%

\section{Introduction}

Understanding neutrino-interactions with nuclei in the broad kinematical region relevant to long-baseline neutrino-oscillation experiments is %an intriguing
a challenging
many-body problem, whose solution requires an accurate and consistent description of the nuclear initial and final states, as well as of the interaction vertex. 
These elements are in fact intimately connected, as the %nuclear 
Hamiltonian determining the nuclear wave functions 
%dictating nuclear interactions from which the correlations originate 
is related to the currents entering the definition of the transition operators through the continuity equation. %Finally
In addition, for large values of the momentum transfer, a full account of relativistic effects and the resonance production mechanism is required. %has to be given. 

The payoff of this endeavor is high, as it will lead to a significant reduction of the systematic uncertainties associated with the determination of oscillation parameters. In addition, a comparison between theoretical predictions and experimental data will provide a great deal of  previously unavailable information, allowing to test the existing models of nuclear interactions and currents, 
notably in kinematical regions sensitive to the high-momentum components of the nuclear wave function. As an example, signatures of nuclear 
short-range correlations arising from the non-central component of  the nucleon-nucleon (NN) force have been recently identified in charge-changing neutrino-nucleus 
interactions observed in the Liquid Argon Time Projection Chamber of the ArgoNeuT Collaboration \cite{Acciarri,Cavanna:2015sla}.

Electroweak currents are usually tested on transitions of light nuclei, characterized by extremely low momentum transfer. Validating these currents in neutrino-nucleus scattering calculations would corroborate their applicability in the lower energy window, down to 20\textendash 30~MeV, which is of great relevance for the physics of supernovae. Finally, probing the high-momentum region is potentially relevant for the ongoing and planned searches of neutrinoless double-beta ($0\nu\beta\beta$) decay. In fact, unlike the standard  $2\nu\beta\beta$ process, in the 
$0\nu\beta\beta$ decay the virtuality of the neutrino in the intermediate state makes the 
%theoretical predictions for the 
nuclear matrix element sensitive to the high momentum components of the nuclear wave function.

The measurement of a Charged Current Quasi Elastic (CCQE) cross section largely exceeding the predictions of the Relativistic Fermi Gas Model (RFGM),  reported by the MiniBooNE collaboration \cite{AguilarArevalo:2007ab,AguilarArevalo:2010zc},  has clearly exposed the need for a more accurate model of neutrino-nucleus interactions, whose development  will  require a cross-disciplinary transfer of knowledge between nuclear theorists, neutrino experimentalists and the
developers of simulation codes. In the pioneering works of Martini {\em et al.}  \cite{Martini:2009uj,Martini:2010ex} and Nieves {\em et al.} \cite{Nieves:2011pp,Nieves:2011yp}, the discrepancy between theory and data has been ascribed to reaction mechanisms other than single nucleon knock out, such as those involving meson-exchange currents (MEC), leading to the occurrence of many-particle many-holes final states. The contributions of MEC, evaluated within the RFGM, have been also included in the phenomenological approach based on the scaling analysis of electron-nucleus scattering data ~\cite{Amaro:2010sd,Megias:2014qva,Megias:2016}. While being remarkably successful in  explaining MiniBooNE data, however, these models are all based on a somewhat oversimplified description of the underlying nuclear dynamics.

%On the other hand, within the super-scaling approach the phenomenological description of the quasielastic region has been supplemented with the inclusion of meson-exchange currents, whose contribution has been evaluated within the relativistic Fermi gas model~\cite{Amaro:2010sd,Megias:2014qva}. Despite their remarkable success, these models are all based on a somewhat oversimplified model of the underlying nuclear dynamics.

Over the past decade, {\em ab initio} approaches have reached the degree of maturity needed to describe lepton-nucleus scattering processes starting from a realistic model of the interactions among the nucleons and between them and the beam particle. For instance, the electric dipole response of $^{16}$O and $^{40}$Ca has been computed combining the Lorentz integral transform with the coupled-cluster many-body technique~\cite{Bacca:2013dma,Bacca:2014rta}. The Green's Function Monte Carlo (GFMC) algorithm has been implemented to perform accurate calculations of the electromagnetic response functions of $^4$He and $^{12}$C in the regime of moderate momentum transfer, which fully include nuclear correlations generated by a state-of-the-art Hamiltonian and consistent meson-exchange currents \cite{Lovato:2015qka,Lovato:2016gkq}. The main drawbacks of this method are its computational 
cost\textemdash $\sim$5 million core-hours to compute the response functions for a single value of the momentum transfer\textemdash and the severe difficulties involved in its extension  to include relativistic kinematic and resonance production. 

At large momentum transfer, the formalism based on spectral function (SF) and factorization of the nuclear transition matrix elements \cite{PRD} allows the combination of a fully relativistic description of the electromagnetic interaction with an accurate treatment of nuclear dynamics. Recently, this approach has been generalized to include the contributions of meson-exchange currents leading to final states with two nucleons in the continuum~\cite{Benhar:2015ula,Rocco:2015cil}.  However, final state interactions (FSI) involving the struck particles are treated as corrections, whose inclusion requires further approximations \cite{Benhar2013,Ankowski2015}

In view of the  the above considerations, a comparison of the results obtained using 
%{\em ab initio} and more phenomenological 
different approaches appears to be  much needed. 
Ab initio methods, while not being best suited to study the kinematical region relevant to neutrino experiments,  can in fact provide strict
benchmarks, valuable to constrain more approximate models in the limit of low momentum transfer. 

This article can be seen as a first step in this direction. We report the results of an analysis of the electromagnetic responses of $^{12}$C, obtained  from the GFMC and
SF approaches in a variety of kinematical setups. Our work is
%report the results of a comparison between the electromagnetic responses of $^{12}$C, obtained  from the 
%SF and the GFMC approaches in a variety of kinematical setups, 
aimed at gauging the accuracy of the factorization approximation and the importance of relativistic effects, in both the  kinematics and the current operator. In order to pin down the role played by the elements of the calculations, we only consider one-body terms in the nuclear current, leaving the discussion of two-body MEC to a forthcoming study. It is very important to realize that our comparison is fully legitimate and meaningful, because the SF and GFMC approaches are based on the same dynamical model,  in which nuclear interactions are described by a realistic phenomenological Hamiltonian.

In Section \ref{sec:nr}, we outline the derivation of  the electromagnetic responses from the electron-nucleus cross section, and discuss the main elements of their 
description within the GFMC and SF approaches.
In Section \ref{section:results} we report the results of our analysis, carried out in the kinematical region corresponding to momentum transfer in the range 300\textendash570 MeV, while in 
Section \ref{conclusions} we summarize our findings and state the conclusions.

\section{Nuclear response}
\label{sec:nr}

The double differential cross section of the inclusive electron-nucleus scattering process in which an electron of initial four-momentum $k_e=(E_e,{\bf k}_e)$ scatters off a nuclear target to a state of four-momentum $k_{e^\prime}=(E_{e^\prime},{\bf k}_{e^\prime})$, the hadronic final state being undetected, can be written in the one-photon-exchange approximation as
\beq
\label{xsec}
 \frac{d^2\sigma}{d E_{e^\prime} d\Omega_{e^\prime}} =\frac{\alpha^2}{q^4}\frac{E_{e^\prime}}{E_e}L_{\mu\nu}W_A^{\mu\nu} \ .
\eeq
In the above equation $\alpha =  1/137$ is the fine structure constant, $d\Omega_{e^\prime}$ is the differential solid angle in the direction specified by the vector ${\bf k}_{e^\prime}$, and $q=k_e - k_{e^\prime} =(\omega,{\bf q})$ is the four momentum transfer. The lepton tensor $L_{\mu\nu}$ is fully specified by the measured electron kinematical variables. The nuclear response is described by the tensor $W_A^{\mu\nu}$, defined as
\begin{align}
\label{response:tensor}
W^{\mu \nu}_A({\bf q},\omega) =&\sum_N \langle  0| J_A^\mu(q) | N \rangle  \langle  N | J_A^\nu(q) |   0 \rangle \times \nonumber \\
& \delta^{(4)}(P_0+q-P_N)   \ ,
\end{align}
where $| 0 \rangle$ and $| N \rangle$ denote the initial and final hadronic states, the four-momenta of which are $P_0\equiv ( E_0,{\bf p}_0 )$ and 
$P_N \equiv (E_N,{\bf p}_N) $.

The target ground state $|0\ra$ does not depend on momentum transfer, and can be safely described using nonrelativistic nuclear many-body theory (NMBT). Within this scheme, the nucleus is viewed as a collection of $A$ pointlike protons and neutrons, whose dynamics are described by the Hamiltonian
\beq
\label{NMBT:ham}
H=\sum_{i}\frac{{\bf p}_i^2}{2m}+\sum_{j>i} v_{ij}+ \sum_{k>j>i}V_{ijk}\ .
\eeq
In the above equation, ${\bf p}_i$ is the momentum of the $i$-th nucleon, while the potentials $v_{ij}$ and $V_{ijk}$ describe
two- and three-nucleon interactions, respectively. Phenomenological two-body potentials are obtained from an accurate fit to the available
data on the two-nucleon system, in both bound and scattering states, and reduce to the Yukawa one-pion-exchange potential at large
distances. In this work, we adopt the state-of-the-art 
%parametrization of $v_{ij}$ adopted in this work is the 
Argonne $v_{18}$ potential \cite{Wiringa:1994wb}.
The inclusion of the additional three-body term, $V_{ijk}$, is needed to explain the binding energies of the
three-nucleon systems and nuclear matter saturation properties~\cite{Pieper:2001ap}. 

The nuclear electromagnetic current is usually written as a sum of one- and two-nucleon contributions according to 
\beq
\label{nuclear:current}
J_A^\mu= \sum_i j^\mu_i+\sum_{j>i} j^\mu_{ij} \ , 
\eeq
where the second term in the right hand side\textemdash accounting for processes in which
the photon couples to a meson exchanged between two interacting nucleons or to the excitation of a resonance (see, e.g., Ref.~\cite{Marcucci:2015rca})\textemdash 
is needed for the continuity equation to be satisfied.

In this paper we will discuss the results obtained retaining only the current $j_i^\mu$, which describes interactions involving a single
nucleon. In the quasi elastic (QE) sector, it can be expressed in terms of the measured proton and neutron vector  form factors \cite{Kelly2004,BBBA}. 

Both the current operator and the final nuclear state $|N\ra$, which includes at least one particle carrying a  momentum of order
$|{\bf q}|$, explicitly depend on ${\bf q}$. As a consequence, in the absence of a comprehensive  relativistic description of nuclear 
structure and dynamics, a consistent theoretical calculation of the response tensor is only possible in the kinematical regime 
corresponding to $|{\bf q}|/m\ll 1$, with $m$ being the nucleon mass, where the non relativistic approximation is applicable.

By performing the Lorentz contraction, the double differential cross section of Eq.\eqref{xsec}, can be written in terms of the nuclear responses describing interactions with longitudinally (L) and transversely (T) polarised photons
\begin{align}
\nonumber
\frac{d^2\sigma}{d E_e^\prime d\Omega_e}&  =\left( \frac{d \sigma}{d \Omega_e} \right)_{\rm{M}} \Big[  A_L(|{\bf q}|,\omega,\theta_e)  R_L(|{\bf q}|,\omega) \\
& + A_T(|{\bf q}|,\omega,\theta_e)  R_T(|{\bf q}|,\omega) \Big] \ ,
\label{xsec:RL:RT}
\end{align}
where 
\begin{align}
A_L = \Big( \frac{q^2}{{\bf q}^2}\Big)^2  \ \ \ , \ \ \ A_T = -\frac{1}{2}\frac{q^2}{{\bf q}^2}+\tan^2\frac{\theta_e}{2}  \ .
\end{align}
and $( d \sigma/d \Omega_e)_{\rm{M}}= [ \alpha \cos(\theta_e/2)/4 E_e\sin^2(\theta_e/2) ]^2$ is the Mott cross section.

The $L$ and $T$ response functions can be readily expressed in terms of the components of the hadron tensor, {\em i.e.} of the nuclear current matrix elements of Eq. \eqref{nuclear:current}, as
\begin{align}
\label{RL}
R_L & = W^{00}_A\nonumber \\
& = \sum_N \langle  0| J_A^0 | N \rangle  \langle  N | J_A^0 |   0 \rangle \delta^{(4)}(P_0+q-P_N)   \ ,\\
\label{RT}
R_T &= \sum_{ij=1}^3\Big(\delta_{ij}-\frac{q_iq_j}{{\bf q}^2}\Big) W^{ij}_A\nonumber\\
&=\sum_N \langle  0| J_A^T | N \rangle  \langle  N | J_A^T |   0 \rangle \delta^{(4)}(P_0+q-P_N)\ .
\end{align}
Choosing the $z$-axis along the direction of the momentum transfer, one finds
\begin{align}
R_T &= W^{xx}_A+W^{yy}_A= \Big[\langle  0| J_A^x | N \rangle  \langle  N | J_A^x |   0 \rangle\nonumber\\
&+\langle  0| J_A^y | N \rangle  \langle  N | J_A^y |   0 \rangle\Big]\delta^{(4)}(P_0+q-P_N)\ .
\end{align}

\subsection{Quantum Monte Carlo}
GFMC is a suitable framework to carry out accurate calculations of a variety of nuclear properties 
in the non relativistic regime (for a recent review of Quantum Monte Carlo methods for nuclear physics see Ref. \cite{Carlson:2014vla}). 

Valuable information on the L and T responses can be obtained from their Laplace transforms, also referred to as Euclidean
responses~\cite{Carlson:1992ga,Carlson:1994zz}, defined as 
\beq
\widetilde{E}_{T,L}({\bf q}, \tau)= \int_{\omega_{\rm{el}}}^\infty \,{d\omega} e^{-\omega \tau}R_{T,L}({\bf q}, \omega)\ .
\eeq
The lower integration limit $\omega_{\rm{el}}= {\bf q}^2/2M_A$, $M_A$ being the mass of the target nucleus, is the threshold of elastic scattering---corresponding to the 
$|N \rangle = |0 \rangle$ term in the sum of Eq. \eqref{response:tensor}---the contribution of which  is excluded.

Within GFMC, the Euclidean responses are evaluated from 
\begin{align}
\nonumber
\widetilde{E}_L({\bf q},\tau) & = \langle 0| \rho^\ast({\bf q}) e^{-(H-E_0)\tau}  \rho({\bf q})|0\rangle \\ 
& -  |\langle 0 | \rho({\bf q}) | 0 \rangle|^2 e^{-\omega_{\rm el} \tau} \ ,
\label{eq:eucL_mat_el}
\end{align}
and 
\begin{align}
\nonumber
\widetilde{E}_T({\bf q},\tau) & = \langle 0| {\bf j}_T^\dagger({\bf q}) e^{-(H-E_0)\tau} {\bf j}_T({\bf q})|0\rangle \\ 
& -  |\langle 0 | {\bf j}_T({\bf q}) | 0 \rangle|^2 e^{-\omega_{\rm el} \tau} \ ,
\label{eq:eucT_mat_el}
\end{align}
where $\rho({\bf q})$ and ${\bf j}_T({\bf q})$ denote non relativistic reductions of the nuclear charge and transverse current operators, respectively \cite{Carlson:2001mp}.
Keeping only the leading relativistic corrections, they can be written as
\begin{align}
\rho_i({\bf q})&= \Big[ \frac{\epsilon_i}{\sqrt{1+Q^2/(4 m^2)}}\nonumber\\
&- i \frac{(2\mu_i-\epsilon_i)}{4 m^2}{\bf q}\cdot ({\bm \sigma}_i \times {\bf p}_i)\Big]\ ,\\
\label{j:trans}
{\bf j}^T_i({\bf q}) &= \Big[ \frac{\epsilon_i}{m}{\bf p}_i^T- i \frac{\mu_i}{2m}{\bf q}\times {\bm \sigma}\Big]\ ,
\end{align}
%The following quantities have been introduced 
with
\begin{align}
\epsilon_i&= G^p_{E}(Q^2)\frac{1}{2}(1+\tau_{z,i})+ G_E^n(Q^2)\frac{1}{2}(1-\tau_{z,i})\ ,\nonumber\\
\mu_i &= G_M^p(Q^2)\frac{1}{2}(1+\tau_{z,i})+ G_M^n(Q^2)\frac{1}{2}(1-\tau_{z,i})\ ,
\label{form:fact}
\end{align}
where $G^{p(n)}_E(Q^2)$  and $G^{p(n)}_M(Q^2)$ are the proton (neutron) electric and magnetic form factors, while $\bm{\sigma}_i$ and $\tau_{z,i}$ are the Pauli 
matrices describing the nucleon spin and the third component of the isospin, respectively. 

Although the states $|N \rangle \neq | 0 \rangle$ do not appear explicitly in Eqs. \eqref{eq:eucL_mat_el} and \eqref{eq:eucT_mat_el}, the Euclidean responses  include the effects of FSI of the particles involved in the electromagnetic interaction, both among themselves and with the spectator nucleons. 

Inverting the Laplace transform to obtain the longitudinal and transverse response functions from their Euclidean counterparts involves non trivial difficulties. However, maximum-entropy techniques, based on bayesian inference arguments~\cite{Bryan:1990,Jarrell:1996rrw}, have been successfully exploited to perform accurate inversions, supplemented by reliable estimates of the theoretical uncertainty~\cite{Lovato:2015qka}. In the case of carbon, particular care has to be devoted to the subtraction of contributions arising from elastic scattering and 
the transitions to the low-lying $2^+$ and $4^+$ states~\cite{Lovato:2016gkq}.

\subsection{Spectral function formalism}
In the kinematical region corresponding to $\lambda~\sim~\pi/|{\bf q}| \ll d$, $d$ being the average NN distance in the
target nucleus, nuclear scattering can be approximated with the incoherent sum of scattering processes involving individual nucleons.
This is the conceptual basis of the Impulse Approximation (IA), which obviously entails neglecting the contribution of the 
two-nucleon current. Under the further assumption that the struck nucleon is decoupled from the spectator particles, the final 
state $|N\ra$  can be written in a factorized form according to
\beq
\label{factorization}
|N\rangle \longrightarrow |{\bf p}^\prime \rangle \otimes |R, {\bf p}_R\rangle \ ,
\eeq
where $|{\bf p}^\prime \rangle$ is the hadronic state produced at the electromagnetic vertex, with momentum ${\bf p}^\prime$, 
and $|R, {\bf p}_R\rangle$ describes the residual system, carrying momentum  ${\bf p}_R$ .

Within the IA, the intrinsic properties of both the target nucleus and the spectator system, which are obviously independent of momentum
transfer, are described in terms of the spectral function, defined as
\beq
\label{pke1}
P({\bf p},E) = \sum_R |\langle R, {\bf p}_R | a_{\bf p} | 0 \rangle |^2 \delta(E+E_0-E_R) \ , 
\eeq
which can be obtained from NMBT. In the above equation,  the operator $a_{\bf p}$ removes a nucleon of 
%$E_R$ is the energy of the $(A-1)$-nucleon residual system, and the 
momentum ${\bf p}$ from the nuclear ground state,  leaving the spectator system with an excitation energy $E$. 

In the QE  sector, the nuclear tensor of Eq.~\eqref{response:tensor} can be written as an integral involving the nuclear SF and the incoherent
sum of the elementary matrix elements of the one-body current between free nucleon states. The resulting expression is
 \begin{align}
W^{\mu\nu}_A =&  \int \,{d^3p \ dE} \ P({\bf p},E) 
\sum_i \langle {\bf p}|j^{\mu}_i |{\bf p+q}\rangle\langle {\bf p+ q}|j^{\nu}_i |{\bf p}\ra\nonumber\\
&\ \ \ \times  \frac{m^2}{E(\bf p)E(|\bf p+q|)}  \ \delta \left[\tilde{\omega}+ E({\bf p}) - E(|{\bf p+q}|)\right] \ ,
\label{pke:had:tens}
\end{align}
%where ${\bf p}^\prime= {\bf p+q}$, as required by momentum conservation,  and 
with
\beq
\tilde{\omega}= \omega + m - E -  E({\bf p})= \omega + M_A - E_R - E({\bf p}) \ .
\eeq
The factors $m^2/(E(\bf p)E(|\bf p+q|))$ have been included to take into account the implicit covariant normalization of quadrispinors of the initial and final nucleons in the matrix element of $j^{\mu}_i$.
The right hand side of Eq.\eqref{pke:had:tens} can be further rewritten in terms of the quantity 
\begin{align}
w^{\mu\nu}_i= \langle {\bf p}|j^{\mu}_i |{\bf p} + {\bf q} & \rangle \langle {\bf p+ q}|j^{\nu}_i |{\bf p}\ra
\nonumber \\
&\times \delta \left[ \tilde{\omega} + E({\bf p}) - E(|{\bf p+q}|) \right] \ , 
\end{align}
% Collecting together the above results, the nuclear response tensor can be  cast in the form
%we finally obtain, for a target nucleus with $Z=A/2$ 
to obtain
\begin{align}
\label{hadN:tens}
W^{\mu\nu}_A= \int \,{d^3p \ dE}  \, & P({\bf p},E)  \frac{m^2}{E(\bf p)E(|\bf p+q|) } \   \nonumber\\
& \times [ Z w_p^{\mu\nu} + (A-Z) w_n^{\mu\nu}]\ ,
\end{align}
$A$ and $Z$ being the target mass number and charge, respectively. \\
Note that $w_{p(n)}^{\mu\nu}$ can be directly related to the tensor describing electron scattering off a \textit{free} proton (neutron), carrying momentum ${\bf p}$, at four momentum transfer $\tilde{q}\equiv (\tilde{\omega},{\bf q})$. 
The effect of nuclear binding is taken into account through the replacement
\beq
q\equiv (\omega,{\bf q})\rightarrow \tilde{q}\equiv (\tilde{\omega},{\bf q})\ ,
\eeq
reflecting the fact that a fraction $\delta\omega$ of the 
energy transfer goes into excitation energy of the spectator system. Therefore, the  elementary scattering process is described as if it took place in free space, but with energy transfer 
$\tilde{\omega}= \omega-\delta\omega$. 

Within the IA, 
%using a realistic carbon Spectral function of Ref. \cite{}, 
the non relativistic expression of the longitudinal response reads
\begin{align}
R_L&= \int dE d{\bf p}\ P({\bf p},E)\Big[ \frac{Z G^p_E(\tilde{Q}^2)+ (A-Z) G^n_E(\tilde{Q}^2)}{1+\tilde{Q}^2/(4 m^2)}\Big]\nonumber\\
&\times\delta \left[ \omega + M_A - E_R - E(|{\bf p+q}|) \right] \theta (|{\bf p+q}|- k_F)
\label{pauli}
\end{align} 
where $E(|{\bf p+q}|)= m + |{\bf p+q}|^2/(2m)$ and $E_R= M_R+ {\bf p}^2/(2 M_R)$ are the energies of the knocked out nucleon and  the recoiling system, whose mass is given by $M_R= M_A- m +E$, respectively.

Note that in the spectral function the state describing the initial nucleon in the interaction vertex is completely
antisymmetrized with respect to the other particles in the target nucleus. On the other hand, in the final state only the antisymmetrization of the spectator system is present, while according to the factorization scheme the
antisymmetrization of the struck nucleon with respect to the spectator particles is disregarded. 
As a consequence, the nuclear initial and final states are not orthogonal to one another.  
In Eq.~\eqref{pauli} Pauli's principle is accounted for by requiring the momentum of the knocked out nucleon to be larger than the nuclear 
Fermi momentum $k_F = 211$ MeV, determined following the procedure described in Ref.~\cite{Ankowski2015}. While this prescription 
is admittedly rather crude, being based on the local Fermi gas model of the nuclear ground state, it has to be kept in mind that the effect of 
Pauli blocking rapidly decreases with increasing momentum transfer, and vanishes altogether at $|{\bf q}| \gsim 2k_F$. 
%However, which for $|\mathbf{q}|=300$ and $380$ MeV 
%leads to a significant reduction of the nuclear response in the low energy-transfer region.

%This has a negligible effect for $|\mathbf{q}|\geq 500$ MeV. However in order to evaluate the nuclear response at low momentum, this effect has to be taken into account. If the the both the initial and the final hadronic state are approximated approximated with a transitionally invariant Fermi gas, the fact that the plane wave describing the struck nucleon has momentum larger than $|{\bf k}_F|$ implies that the factorized state is orthogonal to the initial state. Within the SF formalism, adopting this prescription is the most straightforward, albeit not exact, strategy to perform this orthogonalization, which for $|\mathbf{q}|=300$ and $380$ MeV 
%leads to a significant reduction of the nuclear response in the low energy-transfer region.

Using Eq.~\eqref{j:trans}, we obtain the transverse response
\begin{align}
R_T&= \int dE d{\bf p}\ P({\bf p},E) \big[ Z r_T^p + (A-Z) r_T^n\big]\nonumber\\
&\times\delta\left[ \omega + M_A - E_R - E({\bf p+q})\right]\theta (|{\bf p+q}- {\bf k}_F) \ , 
\end{align}
where 
\begin{align}
r_T^p= \Big[ -{G_E^p}^2(Q^2)\frac{{\bf p}_T^2 }{m^2}+ \frac{{G_M^p}^2(Q^2){\bf q}^2}{2 m^2}\Big]\ ,\nonumber\\
r_T^n= \Big[ -{G_E^n}^2(Q^2)\frac{{\bf p}_T^2 }{m^2}+ \frac{{G_M^n}^2(Q^2){\bf q}^2}{2 m^2}\Big]\ .
\end{align}

The relativistic form of the nuclear responses is written in terms of the one-body current
\begin{align} 
j^\mu_i= \frac{(\epsilon_i+\tau \mu_i)}{(1+\tau)} \gamma^\mu + \frac{(\mu_i-\epsilon_i)}{(1+\tau)}\frac{i\sigma^{\mu\nu}q_\nu}{2m}
\end{align}
where $\tau= \tilde{Q}^2/(4 m^2)$ and $\epsilon_i,\ \mu_i$ are defined in Eq.\eqref{form:fact}. 
In this case, the argument of the energy-conserving $\delta$-function, determining the integration limits of the phase-space integration,  
has to be written using the relativistic expression of the kinetic energies of both the knocked out nucleon and the recoiling spectator  system, {\em i.e.} setting   $E(|{\bf p+q}|)=\sqrt{|{\bf p+q}|^2 + m^2}$ and $E_R= \sqrt{|{\bf p}|^2 + M_R^2}$.

The GFMC and SF approaches consistently account for NN correlations both in the nuclear ground state and among the (A-1) spectator particles. The continuum contribution to the  SF is 
in fact obtained  from the same hamiltonian employed in the GFMC's imaginary time evolution. The main assumption implied in the factorization {\em ansatz} underlying the IA is that FSI between the struck particle and the spectator system, as well as orthogonality  between the initial and final nuclear states,  can be neglected in the limit of high momentum transfer. Because the nuclear response is only sensitive to FSI taking place within a distance $\sim 1/|{\bf q}|$ of the electromagnetic vertex, at high momentum transfer this assumption appears to be largely justified. However, FSI effects  can be sizable at low momentum transfer, and their effect must be consistently taken into account \cite{Benhar2013}. In this work,  we have followed the  scheme developed by the authors of 
Ref.~\cite{Ankowski2015}, which proved very effective in describing FSI in electron-carbon scattering in a broad kinematical range. 

 %%%%%%%%%%%%%%%%%%%%%%%%%%%%%%%%%%%%%%%%%%%%%%%%%%%%%%%%%%
\begin{figure}[h!]
\includegraphics[scale=0.675]{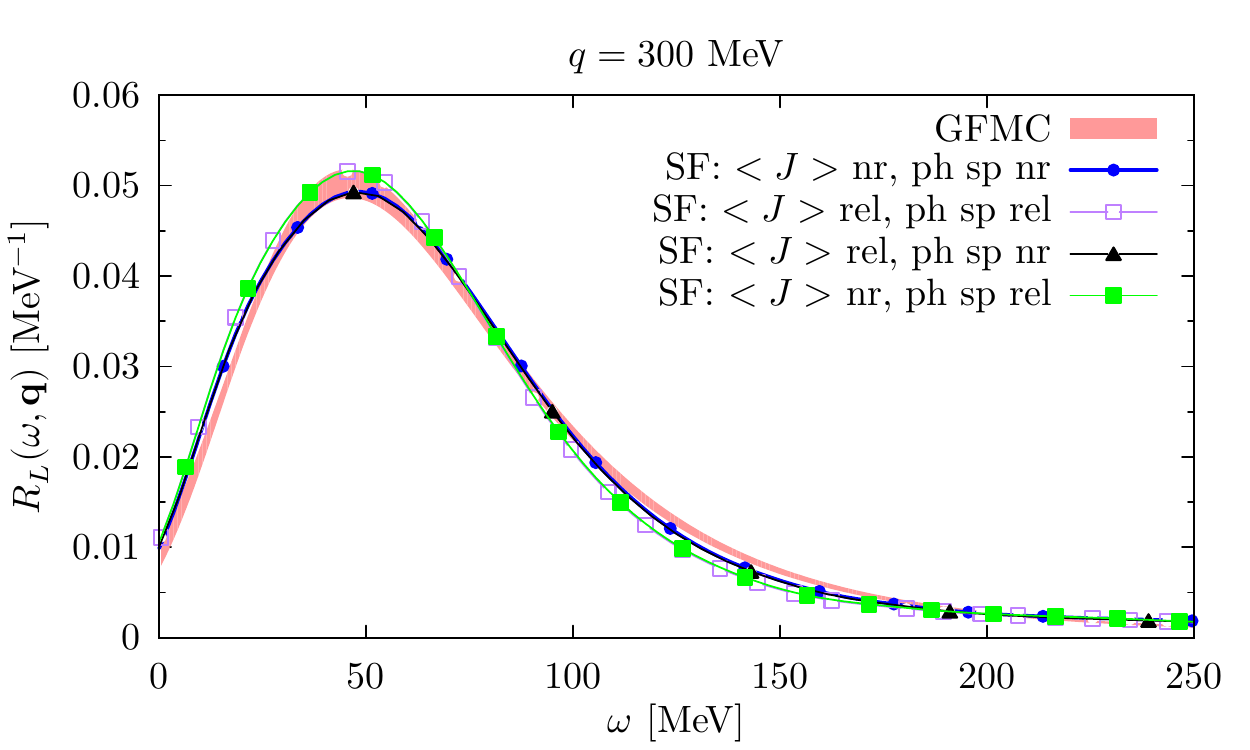}
\vspace*{-.1in}
\caption{Electromagnetic response of Carbon in the longitudinal channel at $|{\bf q}|= 300$ MeV.  The shaded area shows the results of the GFMC calculation, with the associated uncertainty arising from the inversion of the Euclidean response. All the remaing curves have been obtained within the SF approach, including the effects of Pauli blocking and FSI.
The lines marked with dots and hollow squares correspond to non relativistic and fully relativistic calculation, respectively.  Those marked with triangles and filled squares have 
been obtained performing hybrid calculations: non relativistic current and relativistic phase space (triangles), or relativistic current and non relativistic phase space (filled squares).}
\label{300_rl}
\end{figure}
%%%%%%%%%%%%%%%%%%%%%%%%%%%%%%%%%%%%%%%%%%%%%%%%%%%%%%%%%%

 %%%%%%%%%%%%%%%%%%%%%%%%%%%%%%%%%%%%%%%%%%%%%%%%%%%%%%%%%%
\begin{figure}[h!]
\includegraphics[scale=0.675]{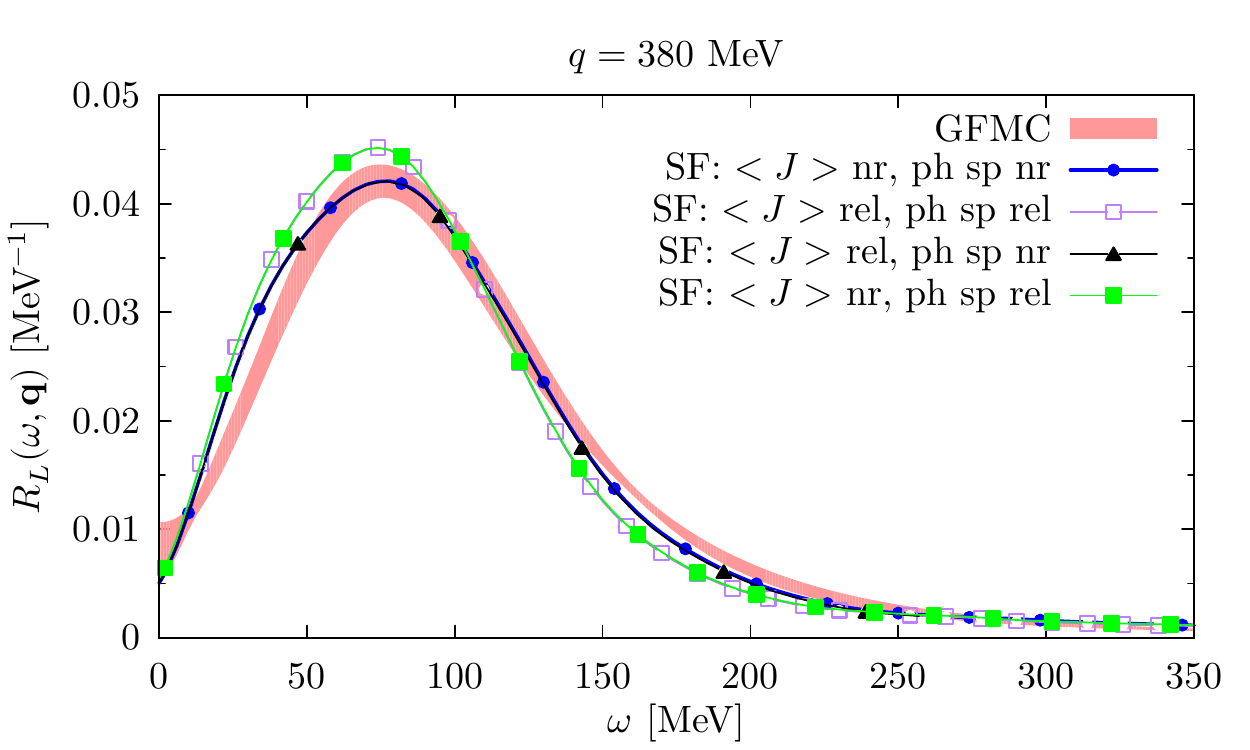}
\vspace*{-.1in}
\caption{ Same as Fig. \ref{300_rl} for $|{\bf q}|= 380$ MeV.}
\label{380_rl}
\end{figure}
%%%%%%%%%%%%%%%%%%%%%%%%%%%%%%%%%%%%%%%%%%%%%%%%%%%%%%%%%%

 %%%%%%%%%%%%%%%%%%%%%%%%%%%%%%%%%%%%%%%%%%%%%%%%%%%%%%%%%%
\begin{figure}[h!]
\includegraphics[scale=0.675]{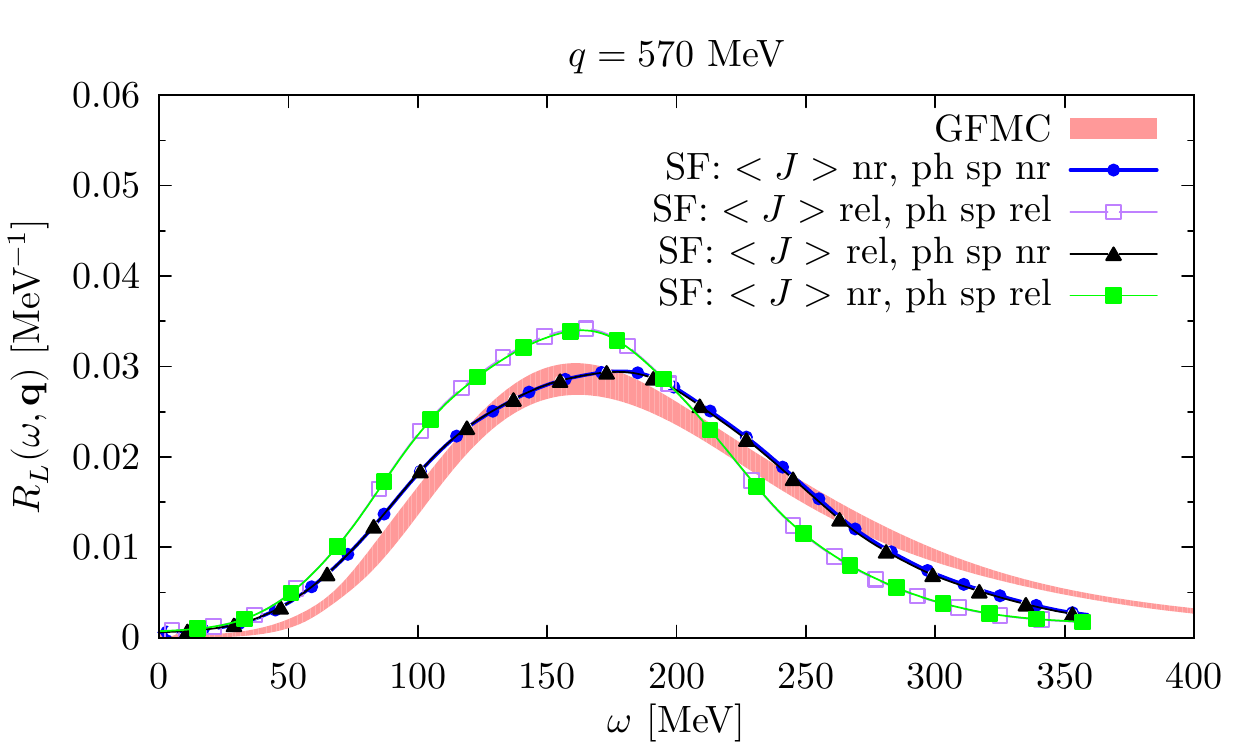}
\vspace*{-.1in}
\caption{Same as Fig. \ref{300_rl} for $|{\bf q}|= 570$ MeV.}
\label{570_rl}
\end{figure}
%%%%%%%%%%%%%%%%%%%%%%%%%%%%%%%%%%%%%%%%%%%%%%%%%%%%%%%%%%

 %%%%%%%%%%%%%%%%%%%%%%%%%%%%%%%%%%%%%%%%%%%%%%%%%%%%%%%%%%
\begin{figure}[h!]
\includegraphics[scale=0.675]{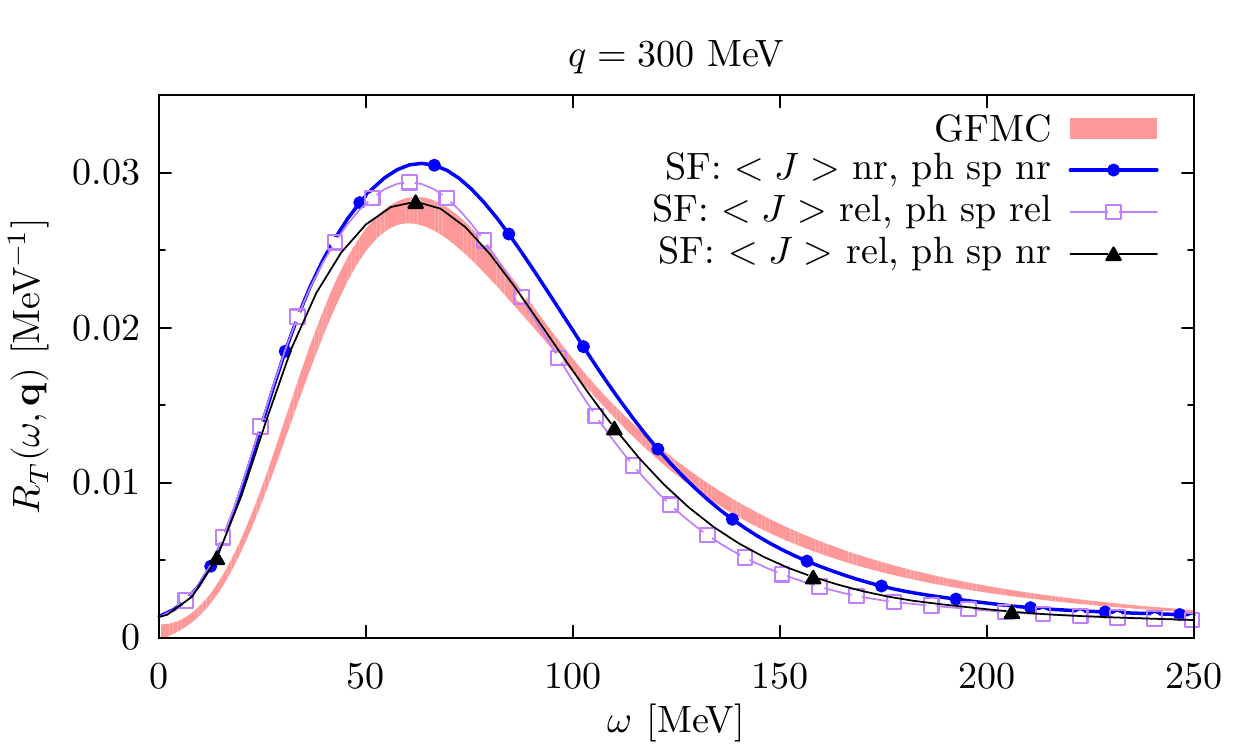}
\vspace*{-.1in}
\caption{Electromagnetic response of Carbon in the transverse channel at $|{\bf q}|= 300$ MeV.  The shaded area shows the results of the GFMC calculation, with the associated uncertainty arising from the inversion of the Euclidean response. All the remaing curves have been obtained within the SF approach, including the effects of Pauli blocking and FSI. 
The lines marked with dots and hollow squares correspond to non relativistic and fully relativistic calculation, respectively, while the one marked with triangles has been obtained performing an hybrid calculation: relativistic current and non relativistic phase space.}
\label{300_rt}
\end{figure}
%%%%%%%%%%%%%%%%%%%%%%%%%%%%%%%%%%%%%%%%%%%%%%%%%%%%%%%%%%

 %%%%%%%%%%%%%%%%%%%%%%%%%%%%%%%%%%%%%%%%%%%%%%%%%%%%%%%%%%
\begin{figure}[h!]
\includegraphics[scale=0.675]{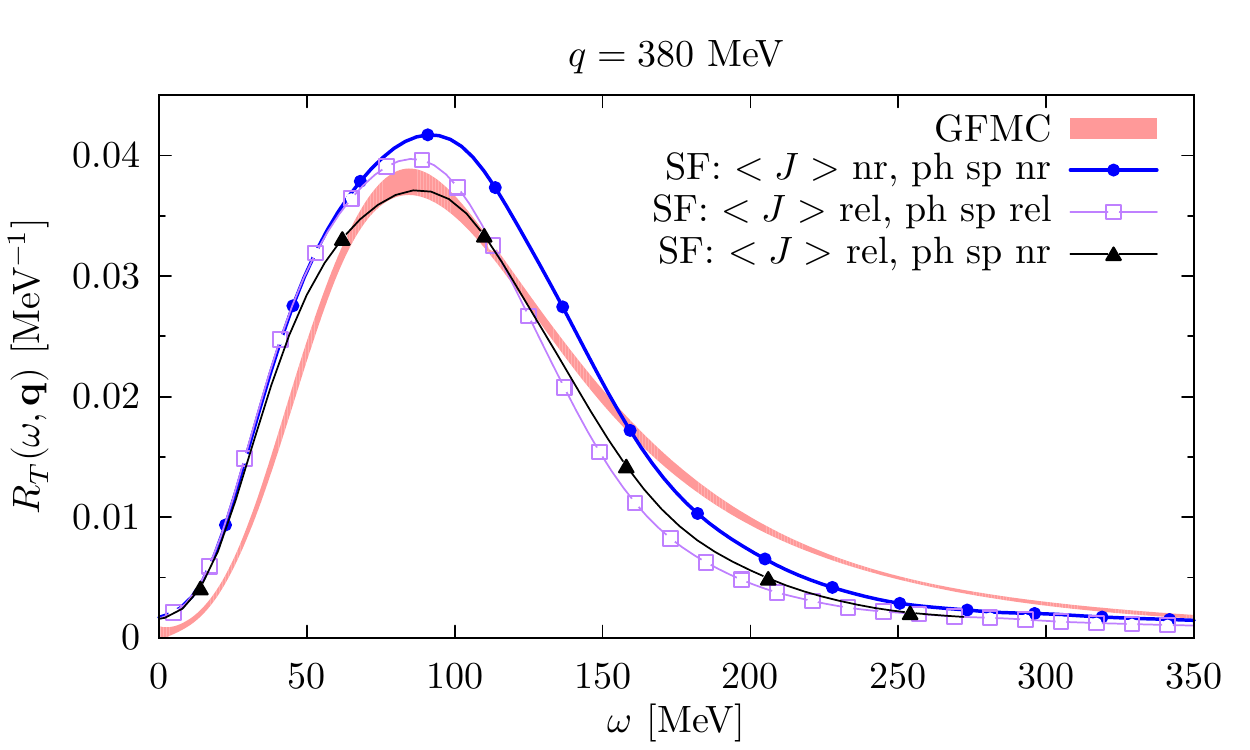}
\vspace*{-.1in}
\caption{Same as Fig. \ref{300_rt} for $|{\bf q}|= 380$ MeV.}
\label{380_rt}
\end{figure}
%%%%%%%%%%%%%%%%%%%%%%%%%%%%%%%%%%%%%%%%%%%%%%%%%%%%%%%%%%

 %%%%%%%%%%%%%%%%%%%%%%%%%%%%%%%%%%%%%%%%%%%%%%%%%%%%%%%%%%
\begin{figure}[h!]
\includegraphics[scale=0.675]{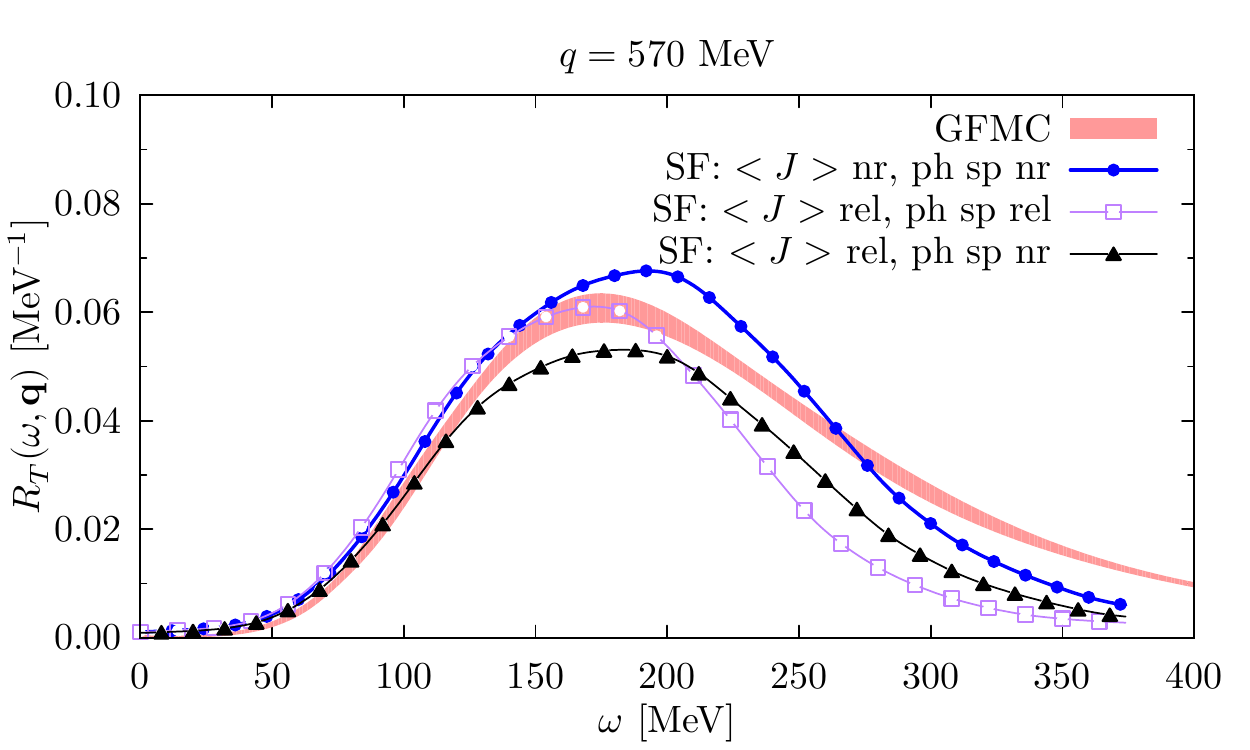}
\vspace*{-.1in}
\caption{Same as Fig. \ref{300_rt} for $|{\bf q}|= 570$ MeV.}
\label{570_rt}
\end{figure}
%%%%%%%%%%%%%%%%%%%%%%%%%%%%%%%%%%%%%%%%%%%%%%%%%%%%%%%%%%
\section{Results}
\label{section:results}

In Figs. \ref{300_rl}, \ref{380_rl}, and \ref{570_rl} we show the results of the GFMC and SF calculations of the electromagnetic responses of Carbon in the longitudinal channel, for momentum transfer $|{\bf q}|= 300,\ 380$ and $570$ MeV. 

Overall, the emerging pattern suggests that\textemdash once Pauli blocking and FSI are accounted for\textemdash the agreement between the two methods is quite good, provided the non relativistic expression for the current operators and the phase space are consistently employed. 

The four different curves labelled SF correspond to the different prescriptions to include relativistic effects.
%adopted for the current operator and the phase space. In particular 
The lines marked with dots and hollow squares represent the non relativistic and fully relativistic results, respectively,  while those marked with triangles and filled squares have been obtained 
performing hybrid calculations, in which the non relativistic expressions have been only used either for the current or for the phase space. 

It is apparent that at $|{\bf q}|= 300$ and $380$ MeV relativistic effects are small. There is little spread between the four SF curves, which are all very close to that 
corresponding to the GFMC calculation.

At $|{\bf q}|= 570$ MeV, SF and GFMC still give very similar results provided the SF calculation is carried out  using relativistic currents and non relativistic phase space. 
On the other hand, the results of the fully relativistic calculation and those obtained using non relativistic currents and relativistic phase space clearly show that 
in this kinematical setup relativistic effects\textemdash comprised in the energy conserving $\delta$-function\textemdash are sizable, and lead to a shift and an enhancement of the peak, 
whose width is reduced.

In Figs. \ref{300_rt}, \ref{380_rt}, and \ref{570_rt} we show the electromagnetic response of Carbon in the transverse channel for the same three values of $|{\bf q}|$.

 The agreement between  the GFMC and the SF results is not as good as in the longitudinal case. Furthermore, the different behaviour of the curves corresponding to the 
three SF calculations deserves some comments. As already pointed out in  the discussion of the longitudinal responses, a comparison between the relativistic and the hybrid calculations 
performed with the non relativistic phase space clearly shows that using relativistic kinetic energies in the argument of the energy-conserving $\delta$-function results in a shift and 
 an enhancement of the peak of the response. However, unlike the longitudinal one, the transverse response is strongly affected by relativistic effects arising from the treatment of the current operator. 
This feature clearly manifests itself in the different shapes exhibited by the results of the non relativistic SF calculations and those of the hybrid calculations performed using relativistic currents and non relativistic kinetic energies.

\section{Conclusions}
\label{conclusions}

The electromagnetic response functions of carbon in the longitudinal and transverse channels have been evaluated within the GFMC and SF approaches at momentum tranfer $|{\bf q}| = 300, \ 380$ and $570$ MeV. 

Because all calculations have been carried out using the same nuclear Hamiltonian and current operator, the resulting response functions can be can be meaningfully compared, to shed light on 
the limits of applicability of both the IA, providing the basis of the SF formalism, and the non relativistic approximation inherent in the GFMC method.

Overall, we find that the GFMC results are in remarkably good agreement with those obtained from the SF approach using non relativistic kinetic energies and currents, provided corrections 
arising from FSI and Pauli blocking are taken into account.

The emerging pattern strongly suggests that the factorization approximation can be safely used down to momentum transfer as low as $\sim 300$ MeV. This is the first important finding of
our study. 

In the longitudinal channel relativistic effects are quite small at momentum transfer 300 and 380 MeV. At $|{\bf q}|~=~570$ MeV they become sizable, and arise mainly from the use of 
relativistic kinetic energies in the argument of the energy-conserving $\delta$-function. The significant reduction of the width can be easily understood considering that its value, while  
increasing linearly with  $|{\bf q}|$ in the non relativistic case, becomes constant and independent of momentum transfer in the relativistic regime.

The interpretation of the results in the transverse channel is more complex. Within the factorization scheme, the main  elements entering the definition of the nuclear response are the nuclear amplitudes and the matrix elements of the nuclear current operators. Hence, the accuracy of the results obtained from this approach depends 
 on the treatment of these two quantities.  
In particular, the degree of complexity of the interaction vertex determines the level of accuracy required in the  calculation of the nuclear spectral function. 

As an example, consider that in the transverse channel %nuclear response 
 the matrix element of the current contains terms 
%which are 
linear in the momentum of the struck particle. However,  since 
%we require 
the spectral function of Ref.~\cite{LDA}, employed to carry out our calculations, is spherically symmetric, they do not contribute to the responses. 
A more accurate treatment of the carbon ground state, taking into account its deformation, would allow to include the contributions arising from these terms.
%However, due to the fact that the ground-state wave function of $^{12}$C is deformed, this assumption is not fully justified. Hence, in the transverse response we are neglecting terms that could be non vanishing once the complex structure of the different nuclear levels of $^{12}$C is fully taken into account.

 Contrary to what is observed in the longitudinal channel, in the transverse responses, relativistic effects are to be ascribed not only to the arguments of the energy-conserving $\delta$-function,  but also  to the treatment of the current operator.  The fact
that for large values of the momentum transfer relativistic corrections to the transverse one-body current are important suggests that improving the non relativistic 
expansion with the inclusion of terms 
$\mathcal{O}[(|{\bf q}|/m)^2]$ may be needed obtain a more accurate prediction of the response. 

The analysis reported in this paper provides valuable and novel information, much needed to reach a better understanding of the description of the nuclear cross section 
obtained from different {\em ab initio} approaches. Our study obviously needs to be completed including the contributions of two-nucleon currents, which are 
known to be important in the transverse channel. Work towards the achievement of this goal is underway.

\begin{acknowledgments}

Many illuminating discussions and a critical reading of the manuscript by Rocco Schiavilla are gratefully acknowledged. NR thanks the Theory Group at TRIUMF for its hospitality and for partial support during the completion of this work. The work of NR has been partially supported by the Spanish Ministerio de
Economia y Competitividad and European FEDER funds under the contracts FIS2014-51948-C2-1-P. The work of OB and NR was supported by INFN under grant MANYBODY. The work of AL was supported by the U.S. Department of Energy, Office of Science, Office of Nuclear Physics, under contract DE-AC02-06CH11357.

\end{acknowledgments}

%{\blue Provided that in the SF approach final state interactions between the struck particle and the spectator system are accounted for, we find very good agreement between GFMC and the SF results, when the nonrelativsitc kinematics and currents operators are employed. }

\bibliography{biblio}

\end{document}